\documentclass[%
 reprint, 
superscriptaddress,
 amsmath,amssymb,
 aps,
]{revtex4-2}

\usepackage{graphicx}
\usepackage{dcolumn}
\usepackage{bm}
\usepackage{ulem}

\usepackage{xcolor}

\graphicspath :
\graphicspath{{Figures/}}

\begin{document}

\preprint{APS/123-QED}

\title{Loss of coherence and change in emission physics \\for radio emission from very inclined  cosmic-ray air showers}

\author{Simon Chiche}
 \email{simon.chiche@iap.fr}
\affiliation{%
 Sorbonne Universit\'{e}, CNRS, UMR 7095, Institut d'Astrophysique de Paris, 98 bis bd Arago, 75014 Paris, France}\
%
\author{Chao Zhang}%
\email{2006385407@pku.edu.cn}
\affiliation{Department of Astronomy, School of Physics, Peking University, Beijing 100871, China}\
\affiliation{Kavli Institute for Astronomy and Astrophysics, Peking University, Beijing 100871, China}\
\affiliation{Institute for Astroparticle Physics (IAP), Karlsruhe Institute of Technology, Karlsruhe, Germany
}%
\affiliation{Sorbonne Universit\'{e}, Universit\'{e} Paris Diderot, Sorbonne Paris Cit\'{e}, CNRS, Laboratoire de Physique Nucl\'{e}aire et de Hautes Energies (LPNHE), 4 place Jussieu, F-75252, Paris Cedex 5, France}\

\author{Felix Schlüter}
\affiliation{Inter-University Institute For High Energies (IIHE), Université libre de Bruxelles (ULB), Boulevard du Triomphe 2, 1050 Brussels, Belgium}

\author{\\Kumiko Kotera}
\affiliation{%
 Sorbonne Universit\'{e}, CNRS, UMR 7095, Institut d'Astrophysique de Paris, 98 bis bd Arago, 75014 Paris, France}\
\affiliation{
Astrophysical Institute, Vrije Universiteit Brussels, Pleinlaan 2, 1050 Brussels, Belgium
}%

\author{Tim Huege}
\affiliation{Institute for Astroparticle Physics (IAP), Karlsruhe Institute of Technology, Karlsruhe, Germany
}%
\affiliation{
Astrophysical Institute, Vrije Universiteit Brussels, Pleinlaan 2, 1050 Brussels, Belgium
}%

\author{Krijn D. de Vries}
\affiliation{
Vrije Universiteit Brussel (VUB), Dienst ELEM, Pleinlaan 2, B-1050, Brussels, Belgium
}%
\author{Matias Tueros}
\affiliation{
IFLP - CCT La Plata - CONICET, Casilla de Correo 727 (1900) La Plata, Argentina}%
\affiliation{Depto. de Fisica, Fac. de Cs. Ex., Universidad Nacional de La Plata,  Casilla de Coreo 67 (1900) La Plata, Argentina
}%
\affiliation{%
 Sorbonne Universit\'{e}, CNRS, UMR 7095, Institut d'Astrophysique de Paris, 98 bis bd Arago, 75014 Paris, France}\
\author{Marion Guelfand}
\affiliation{Sorbonne Universit\'{e}, Universit\'{e} Paris Diderot, Sorbonne Paris Cit\'{e}, CNRS, Laboratoire de Physique Nucl\'{e}aire et de Hautes Energies (LPNHE), 4 place Jussieu, F-75252, Paris Cedex 5, France}\
\affiliation{%
 Sorbonne Universit\'{e}, CNRS, UMR 7095, Institut d'Astrophysique de Paris, 98 bis bd Arago, 75014 Paris, France}\


\bibliographystyle{apsrev4-1}
\begin{abstract}
 Next-generation radio experiments such as the Radio Detector of the upgraded Pierre Auger Observatory and the planned GRAND and BEACON arrays target the detection of ultra-high-energy particle air showers
arriving at low elevation angles.
   These inclined cosmic-ray air showers develop higher in the atmosphere than vertical ones, enhancing magnetic deflections of electrons and positrons inside the cascade. 
   We evidence two novel features in their radio emission: 
   a new polarization pattern, consistent with a geo-synchrotron emission model and a coherence loss of the radio emission, both for showers with zenith angle $\theta \gtrsim 65^{\circ}$ and strong enough magnetic field amplitude (typical strength of $B\sim 50\, \rm \mu T$). Our model is compared with both ZHAireS and CoREAS Monte-Carlo simulations. Our results break the cannonical description of a radio signal made of Askaryan and transverse current emission only, and 
 provide guidelines for the detection and reconstruction strategies of next-generation experiments, including cosmic-ray/neutrino discrimination.



\end{abstract}

\maketitle


\section{\label{sec:intro}Introduction\protect\\}

Radio detection of air showers is a robust and reliable technique to probe the low fluxes of astroparticles at ultra-high energies \cite{Huege2016,Schr_der_2017}. 
The radio signal from near {\it vertical} air showers, induced by primary particles arriving with zenith angles up to $\sim 60^{\circ}$, has been extensively described in the past decade, with theoretical descriptions matching numerical simulations and experimental data \cite{Aab_2016, Buitink_2016}. Yet, next-generation experiments such as AugerPrime Radio~\cite{Pont:2021pwd, Aprime}, BEACON~\cite{Beacon}, and GRAND~\cite{_lvarez_Mu_iz_2019}, now target air showers with {\it very inclined} arrival directions with zenith angles $\theta >65^\circ$. The main reason for this is to increase their sensitivity, exploiting the large radio-emission footprints of these showers~\cite{Huege2016,Aab_2018}.  

In this work, we demonstrate that the existing theoretical descriptions require to be broadened to enfold the case of inclined air showers, resulting in a new paradigm for the nature of radio emission. 


\section{\label{sec:paradigm}New paradigm for radio emission \\and associated signatures\protect\\}

When an ultra-high-energy cosmic ray enters the atmosphere, it produces an air shower, i.e., a cascade of high energy particles. At energies $\epsilon_e>88\,$MeV, electrons and positrons in the cascade undergo  significant Bremsstrahlung  radiation, with attenuation length $l_{\rm rad} = X_0/\rho_{\rm air} = 3.67\times 10^2\,{\rm m}\,(\rho_{\rm air}/1\,{\rm kg\,m^{-3}})^{-1}$, where $X_0=36.7\,{\rm g}\,{\rm cm}^{-2}$ is the electron radiation length. 

In the commonly adopted paradigm of radio emission by air showers, 
the major process, called ``geomagnetic emission" or ``transverse current" emission, is produced by electrons and positrons drifting laterally at the front of the shower, through the competing effects of magnetic deflection and particle scattering through elastic collisions with air-molecules, and hence producing a time-varying transverse current 
\cite{Scholten_2008}. 
A second process, 
the ``Askaryan" mechanism, in which electrons from the air atoms accumulate in the shower front, creating a net negative charge excess in the shower plane, 
was shown to represent $\sim 10\%$ of the total radio emission for vertical air showers, and less than a few \% for very inclined air showers \cite{Chiche_2022, Aab_2014, 2023FelixJCAP...01..008S}.


\subsection{\label{sec:char}Characteristics of very inclined air showers}
\vspace{-0.1cm}
In this study, we define very inclined air showers as those induced by primary particles arriving with zenith angles $\theta>65^\circ$. 
Cosmic-ray air showers reach their maximum longitudinal development around an atmospheric depth of $X_{\rm max}\sim 650-750\,{\rm g\,cm^{-2}}$. 
We assume that the air density decreases with altitude following the Linsley model \cite{Linsley}, with a 5-layer exponential dependency, while accounting for the Earth curvature.
 Very inclined cosmic-ray showers develop in the low-density upper layers of the atmosphere $\gtrsim 10 \rm km$ above sea level. 
 They propagate over longer distances than near-vertical ones.  In the following, the air density $\rho_{\rm air}(\theta)$ is evaluated at the location of the maximum shower development $X_{\rm max}$, at a  geometrical distance $d_{\rm obs}$ away from the impact point of the cascade on the ground.
\\

\begin{figure*}[tb]
\centering 
\includegraphics[width=0.95\columnwidth]
{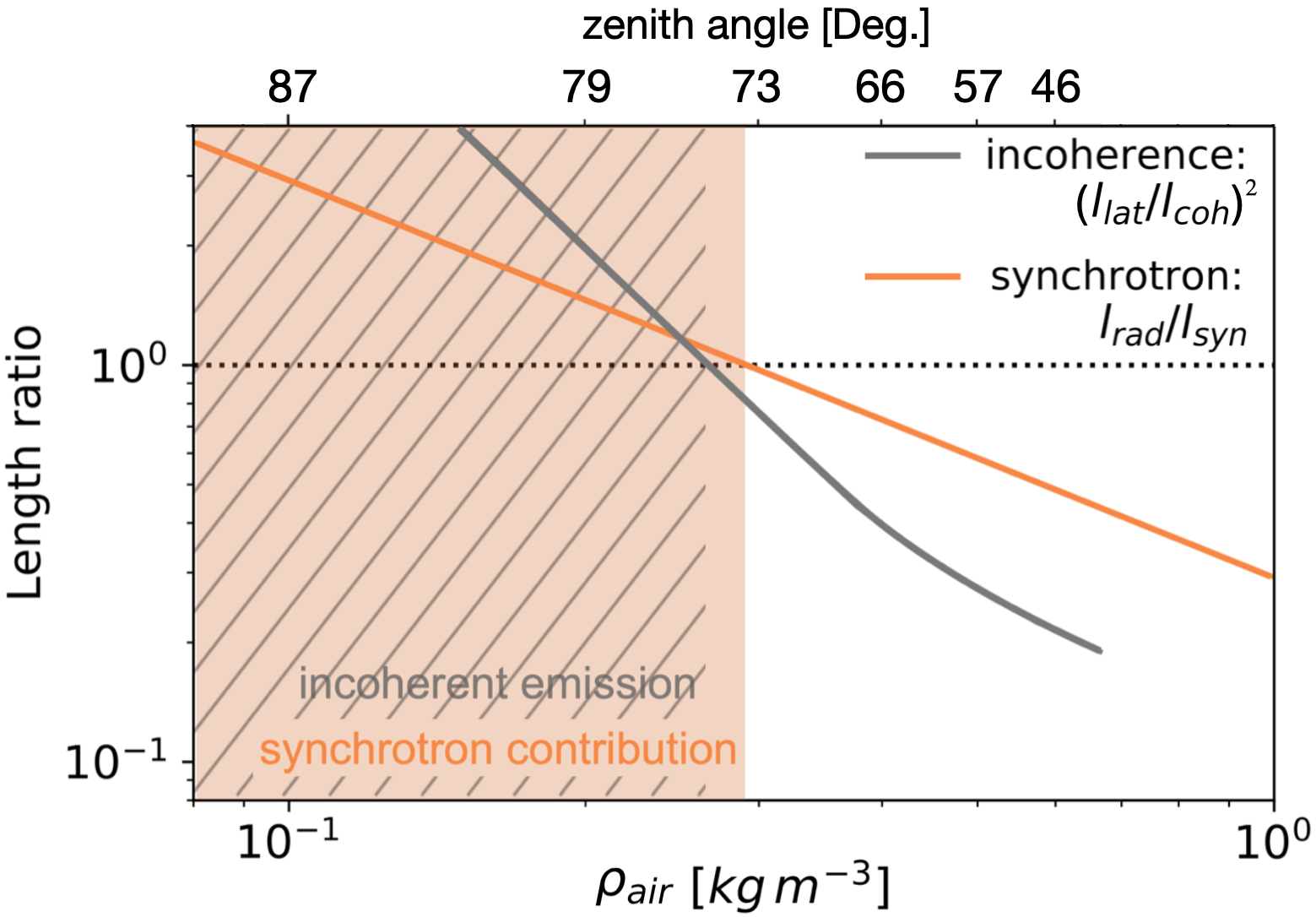}
    \raisebox{-0.03cm}{\includegraphics[width=0.95\columnwidth]{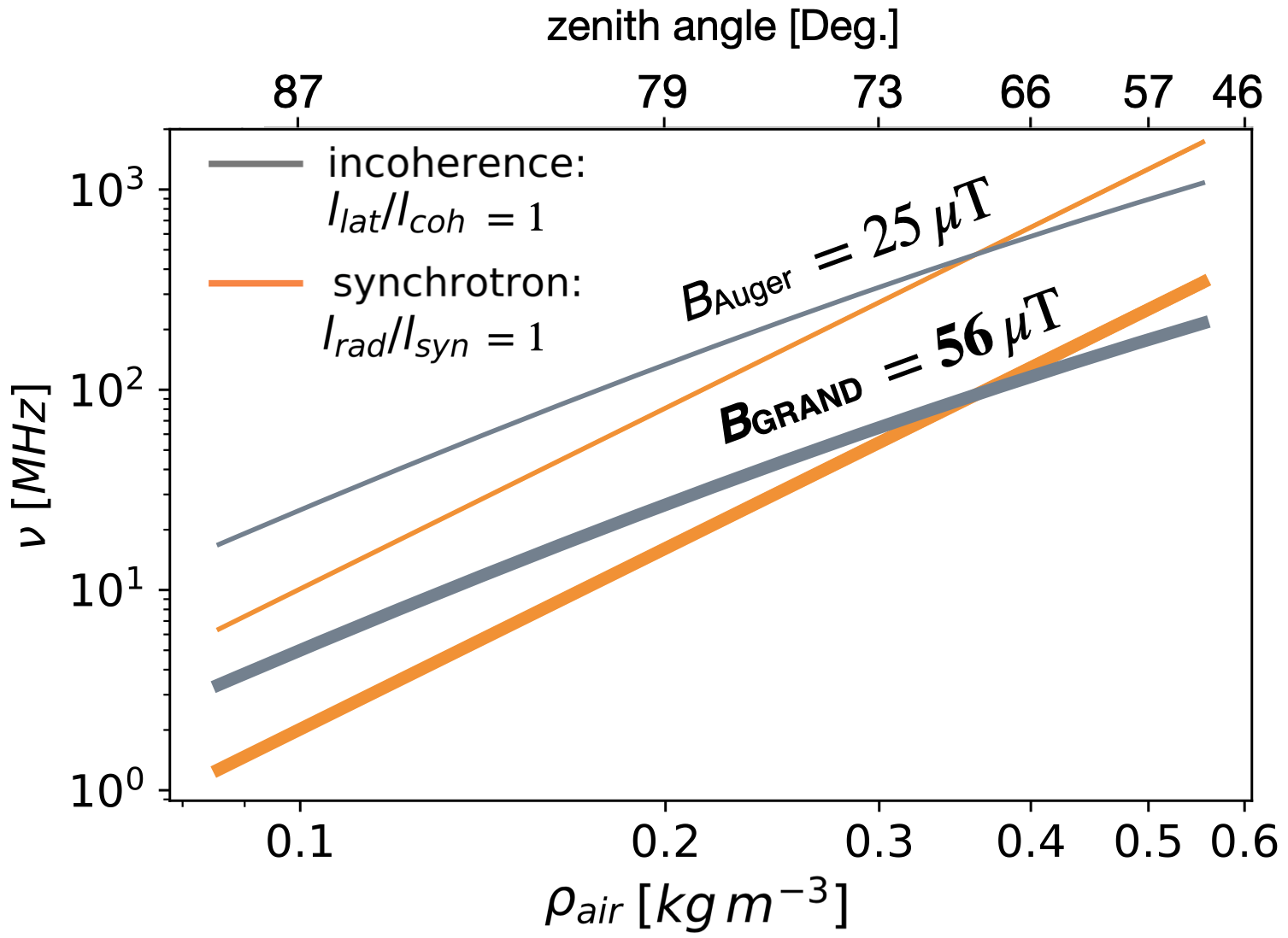}}\caption{ {\it Left:} 
Synchrotron ratio $l_{\rm rad}/l_{\rm syn}$ (orange line) and incoherence ratio $l_{\rm lat}/l_{\rm coh}$ (grey line), as a function of air density at $50\,$MHz, for GRAND site Dunhuang magnetic field strength of $56\, \mu \rm T$. 
The dotted line indicates the limit where the ratios are equal to 1. Both ratios increase with decreasing air density and the transition to a regime where geo-synchrotron emission is expected (predicted in~\cite{James_2022}) is reached for $\rho_{\rm air}< 0.29 \, {\rm kg\,m^{-3}}$ ($\theta>73^\circ$), while the transition to incoherent radio emission is expected for $\rho_{\rm air}< 0.27 \, {\rm kg\,m^{-3}}$ ($\theta>75^\circ$). {\it Right:}  Incoherence and synchrotron transitions (gray: $l_{\rm lat}/l_{\rm coh} =1$ and orange: $l_{\rm syn}/l_{\rm rad} =1$)
in the frequency $\nu$ and air density $\rho$ parameter space, for GRAND (thick lines) and Auger (thin lines) magnetic field strengths. For each case, the region below the gray [orange] line, is a region where a coherent [without significant  geo-synchrotron] emission is expected, while above the line, an incoherent [with significant  geo-synchrotron] emission is expected.} 

\label{fig:regimes}
\end{figure*}

\vspace{-0.8cm}

\subsection{\label{sec:synch}Change in emission mechanisms\protect\\}
\vspace{-0.1cm}

While the term ``synchrotron emission'' canonically refers to the radiation emitted by charges moving on periodic helical trajectories due to magnetic deflections, the term ``geo-synchrotron radiation'' as defined in previous literature~\cite{Falcke_2003,Huege_2005} refers to pairs of $e^{+}$ and $e^{-}$ gyrating for a fraction of a full circle. 
Still, many of the characteristics of classical synchrotron emission, such as the spectral shape at high frequencies and the cutoff at a critical frequency also apply to geo-synchrotron emission from particles traversing incomplete arcs~\cite{Endpoint}.

For vertical air showers, geo-synchrotron emission was shown with simulations and in data to have a negligible contribution, except in the GHz regime~\cite{Huege_James,Smida}.
Indeed, due to multiple Coulomb scattering, a net friction force acts on the particles, which then
diffuse along the transverse direction instead of gyrating on arcs and radiating 
a forward-beamed synchrotron pulse.
In the following, we conservatively estimate the regime where geo-synchrotron emission is allowed, assuming a free synchrotron path ignoring multiple Coulomb scattering.
 
 A transition from transverse currents to geo-synchrotron emission was predicted in~\cite{James_2022} if particles radiate before undergoing Bremsstrahlung interactions, i.e., for  $l_{\rm syn}/l_{\rm rad}<1$, where $l_{\rm syn}\sim 1353 \,{\rm m}\,(\epsilon_e/88\,{\rm MeV})^{\frac{2}{3}}(B\sin{\alpha}/50\,{\rm \mu T})^{-\frac{2}{3}}(\nu/50 \, \rm MHz)^{-\frac{1}{3}}$ \cite{James_2022} is the synchrotron cooling length,  with $B$ the geomagnetic field and $\alpha$  the 
 geomagnetic angle between the shower arrival direction and the local geomagnetic field. We adopt the same formalism but rather assume that the condition $l_{\rm syn}/l_{\rm rad}<1$ indicates the regime where a non-negligible geo-synchrotron component is allowed in addition to the transverse current emission.  From this, one derives
\begin{eqnarray}    
\frac{l_{\rm syn}}{l_{\rm rad}} &\sim& 3.7\left(\frac{\epsilon_e}{88\,{\rm MeV}}\right)^{\!\frac{2}{3}}\!\left(\frac{50\,{\rm \mu T}}{B\sin{\alpha}}\right)^{\!\!\frac{2}{3}}\!\!\left(\frac{50 \, \rm MHz}{\nu} \right)^{\!\!\frac{1}{3}}\!\!\nonumber\\ 
&&\times\left(\frac{\rho}{1\, \rm kg/m^{3}}\right) \ . \label{eq:synchrotron}
\end{eqnarray}
In Figure~\ref{fig:regimes} ({\it left}), the orange line represents the variation of the ratio $l_{\rm syn}/l_{\rm rad}$ as a function of $\rho_{\rm air}$ and $\theta$, assuming $X_{\rm max} = 650\, \rm g\, cm^{-2}$ and taking
the density at $X_{\rm max}$  (Section~\ref{sec:char}),  for typical values of $B= 56\,\rm \mu T$, $\nu = 50\, \rm MHz$, $\sin{\alpha} =1$ at the critical energy $\epsilon_e= 88\, \rm MeV$. We find that a geo-synchrotron component can contribute to the emission of inclined air showers, i.e., $l_{\rm syn}/l_{\rm rad} <1$, for $\rho_{\rm air}\lesssim 0.29 \, {\rm kg\,m^{-3}}$ ($\theta\gtrsim 73^\circ$).

In Figure~\ref{fig:regimes} ({\it right}), 
 the orange thick [thin] line corresponds to $l_{\rm syn}/l_{\rm rad} = 1$ in the ($\nu$, $\rho$) phase space, for magnetic field strengths $B = 56\,\mu$T [$B = 25\,\mu$T] corresponding to GRAND [Auger] sites respectively. The orange line delimits the regions where the geo-synchrotron component can contribute to the radio emission (above the line) or not (below the line).
Consistently to 
Eq.~(\ref{eq:synchrotron}), it appears that for showers with $\rho_{\rm air} < 0.37 \, {\rm kg\,m^{-3}}$ ($\theta>68^{\circ}$) and GRAND magnetic field strength, a geo-synchrotron contribution is allowed starting from frequencies of tens of MHz and above. 
For more vertical showers, $\rho_{\rm air} > 0.37 \, {\rm kg\,m^{-3}}$($\theta<68^{\circ})$, frequencies of at least hundreds of $\rm MHz$ are requiredto allow for a geo-synchrotron contribution. On the other hand, for the lower amplitude magnetic field of the Auger site, geo-synchrotron emission is only allowed at higher frequencies of hundreds of MHz, even for showers
with $\rho_{\rm air} < 0.2 \, {\rm kg\,m^{-3}}$ ($\theta \sim 80^{\circ}$). 

The polarization pattern expected from such a geo-synchrotron emission was already described in earlier work, which predicted the emergence of a ``clover-leaf" pattern for the $\mathbf{v} \times  (\mathbf{v} \times \mathbf{B})$ polarization component of vertical showers at $\rm GHz$ frequencies~\cite{Huege_James,Huege_2005}  ($\mathbf{v}$ and $\mathbf{B}$ being the shower axis and the local magnetic field direction, respectively). Detecting this feature with future experiments would be a support to our geo-synchrotron description.

 Next to a synchrotron interpretation, a direct consequence of the deflection of electrons and positrons in Earth's magnetic field is that from a macroscopic point of view, a moving dipole is induced~\cite{Scholten_2008}, that could lead to a similar polarization pattern. Both, geo-synchrotron and dipole components should become more prominent moving to less dense air. We, however, do not provide an estimate on the relative strength of the synchrotron component compared to the dipole emission, which is to be investigated in future works.

\subsection{\label{sec:Coherence}Coherence loss\protect\\}

The radio emission of air showers is dominated by the superposition of individual emissions by particles in the shower around the point of maximal longitudinal development $X_{\rm max}$. If all particles in the shower emit in phase, the emissions will interfere constructively ({\it coherent} signal). Alternatively, 
the emitted power will sum up and the resulting {\it incoherent} signal is expected to be weaker by a factor of $\sqrt{N}$, where $N$ is the number of radiators. Coherence for radio emission from a high-energy particle cascade is given by its longest projected length scale. Due to Cerenkov effects~\cite{Krijn2011, de_Vries_2012}, the shower longitudinal profile is strongly compressed and the projected shape of the cascade front becomes the leading scale.
To study the radio signal coherence, we assume that the emission between the center of the shower plane and a position at a lateral distance $l_{\rm lat}$ from $X_{\rm max}$ is coherent if their path length difference to the observer is below half a wavelength, i.e., for $\delta<\lambda/2$ where $\delta \sim l_{\rm lat}^{2}/2{d_{\rm obs}}$, with $d_{\rm obs}$ the distance to the observer. From this, we derive the spatial coherence length $l_{\rm coh}$, which quantifies the largest extent over which the radio signal is coherent. Setting $\delta = \lambda/2$, we get $l_{\rm coh} =  \sqrt{({c}/{\nu}) {d_{\rm obs}}}$
where $\nu$ is the frequency of the radio signal.

Inclined showers 
propagate in a less dense medium than near-vertical showers, resulting in a longer mean free path of collision of deflected particles and a larger shower lateral extent. 
To study the radio signal coherence dependency with zenith angle we extract $d_{\rm obs}(\theta)$ from Monte-Carlo simulations (Section~\ref{sec:char}). The shower lateral extent is computed using the formalism of Ref.~\cite{Scholten_2008}. Taking into account magnetic deflection and interaction with air, the transverse particle acceleration in the shower front at time $t$ is expressed as ${\rm d}^2x_{\rm t}/{\rm d}t^{2}=c^3eB\sin{\alpha}/[\epsilon_e\, \exp(-t/\tau)]$, where $x_{\rm t}$ is the particle transverse position (orthogonal to the shower axis) and $\tau = l_{\rm rad}/c$ the Bremsstrahlung energy loss timescale \cite{Scholten_2008}. 
It yields $x_{\rm t}(t) = \tau^{2}c^{3}eB\sin{\alpha}\, (e^{t/\tau} -1 -{t}/{\tau})/\epsilon_e$. The shower lateral extent is then expressed as $l_{\rm lat} = 2 x_{t}(t=\tau)$, where the factor 2 accounts for the dynamics of positrons and electrons. 
 This derivation leads to a lateral extent $l_{\rm lat} \propto \rho^{-2},\, B$, which was confirmed by CORSIKA simulations for inclined air showers with zenith angle $\theta \gtrsim 65^{\circ}$~\cite{Guelfand2023arXiv231019612G}.

The radio signal is expected to be coherent if the shower lateral extent is smaller than the coherence length, i.e., for $l_{\rm lat}/l_{\rm coh} < 1$, which implies $(l_{\rm lat}/l_{\rm coh})^{2} < 1$, and incoherent otherwise. From our formalism this yields: 
\begin{eqnarray}
\left(\frac{l_{\rm lat}}{l_{\rm coh}} \right)^{2} & =& \frac{\nu \, l_{\rm lat}^{2}}{c\, d_{\rm obs}} \sim 0.018 \,
\left( \frac{\nu}{50\,{\rm MHz}}\right)\left( \frac{B\sin{\alpha}}{50\,{\rm \mu T}}\right)^{2}\nonumber\\ 
&&\times\left( \frac{\epsilon_e}{88\,{\rm MeV}}\right)^{-2}\left( \frac{\rho}{1\,{\rm kg\,m^{-3}}}\right)^{-4}\left(\frac{d_{\rm obs}[\rho]}{10 \rm \, km}\right)^{-1} . \label{eq:coherence}
\end{eqnarray}

On the left-hand panel of Figure~\ref{fig:regimes}, the gray line corresponds to the ratio $(l_{\rm lat}/l_{\rm coh})^{2}$ and  defines the coherent (when the ratio is above the dotted line) and incoherent (ratio below the dotted line) regimes as a function of air shower density,
for typical particle energy at $X_{\rm max}$: $\mathcal{E}_{0} = 88\, \rm MeV$, $B =56 \,\mu \rm T$, $\nu = 50 \, \rm MHz$ and $\sin{\alpha} =1$. It shows that $l_{\rm lat}/l_{\rm coh} >1$, i.e., the emission is incoherent, for inclined air showers with $\rho_{\rm air} < 0.27 \, {\rm kg\,m^{-3}}$ ($\theta >75^\circ$), while  $l_{\rm lat}/l_{\rm coh} <1$ for more vertical showers. 
In the right-hand panel of Figure~\ref{fig:regimes}, we also represent with a gray thick line the transition from  coherent (below the line) to incoherent emission (above the line) in the ($\nu$, $\rho$) phase space, for the GRAND magnetic field configuration. Similarly to the geo-synchrotron case, we find that the transition to incoherent emission is expected at low frequencies for inclined showers and high frequencies for  near-vertical ones. Auger's 
lower magnetic field strength (thin gray line, $B=25\, \rm \mu T$) 
 simply shifts the transition 
to higher frequencies. This shows that the radio signal is expected to remain coherent for inclined showers detected at Auger, but not necessarily at locations with higher magnetic field amplitude.

Such a coherence loss 
will decrease the magnitude of the radio signal. 
This attenuation can be modeled, assuming that the total electric field $\mathcal{E}_{\rm tot}$ measured at the observer location $x$ at time $t$, is given by the sum of the individual contributions from $N$ particles located in a plane perpendicular to the shower axis at $X_{\rm max}$:
\begin{equation}\label{eq:Etot}
     \mathcal{E}_{\rm tot} = \sum_{i=0}^{N}\mathcal{E}_i\cos{(k_i x-\omega_{i}t + \varphi_i)} \, ,
\end{equation}
with $k_i$, $\omega_i$ and $\phi_i$ the wave number, angular frequency and phase angle of the $i^{\rm th}$ particle. One can then derive the radiation energy by averaging the squared electric field over time,  $E_{\rm rad}^{\rm obs} = \left<|\mathcal{E}_{\rm tot}|^{2}\right>$. Assuming that all $N$ particles at $X_{\rm max}$ emit in phase at the same wave number, and modeling the distribution of the energy content in a plane perpendicular to the shower axis with a top-hat distribution along a line, we derive an analytical formula (see Appendix~\ref{app:rad_energy}) which predicts an attenuation of the observed radiation energy for inclined air showers and the transition from a regime where $E_{\rm rad}^{\rm obs} \propto N^{2}$ (coherent emission) to a regime where  $E_{\rm rad}^{\rm obs} \propto N$ (randomly distributed phases), when decreasing the air density. This feature could also be observed by future experiments and such a detection would validate our model of the coherence loss.

\begin{figure*}
\centering 
\includegraphics[width=0.95\columnwidth]{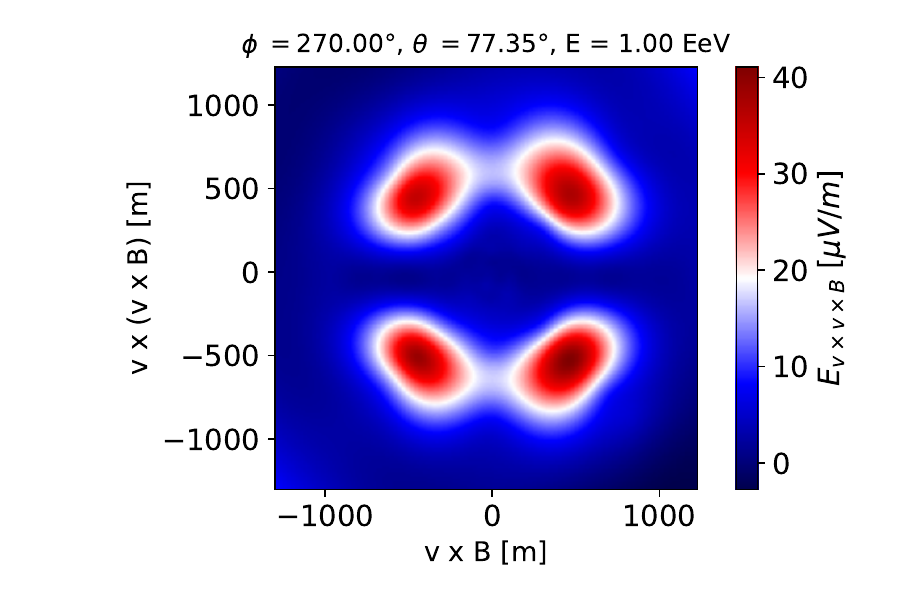}
\includegraphics[width=1.05\columnwidth]{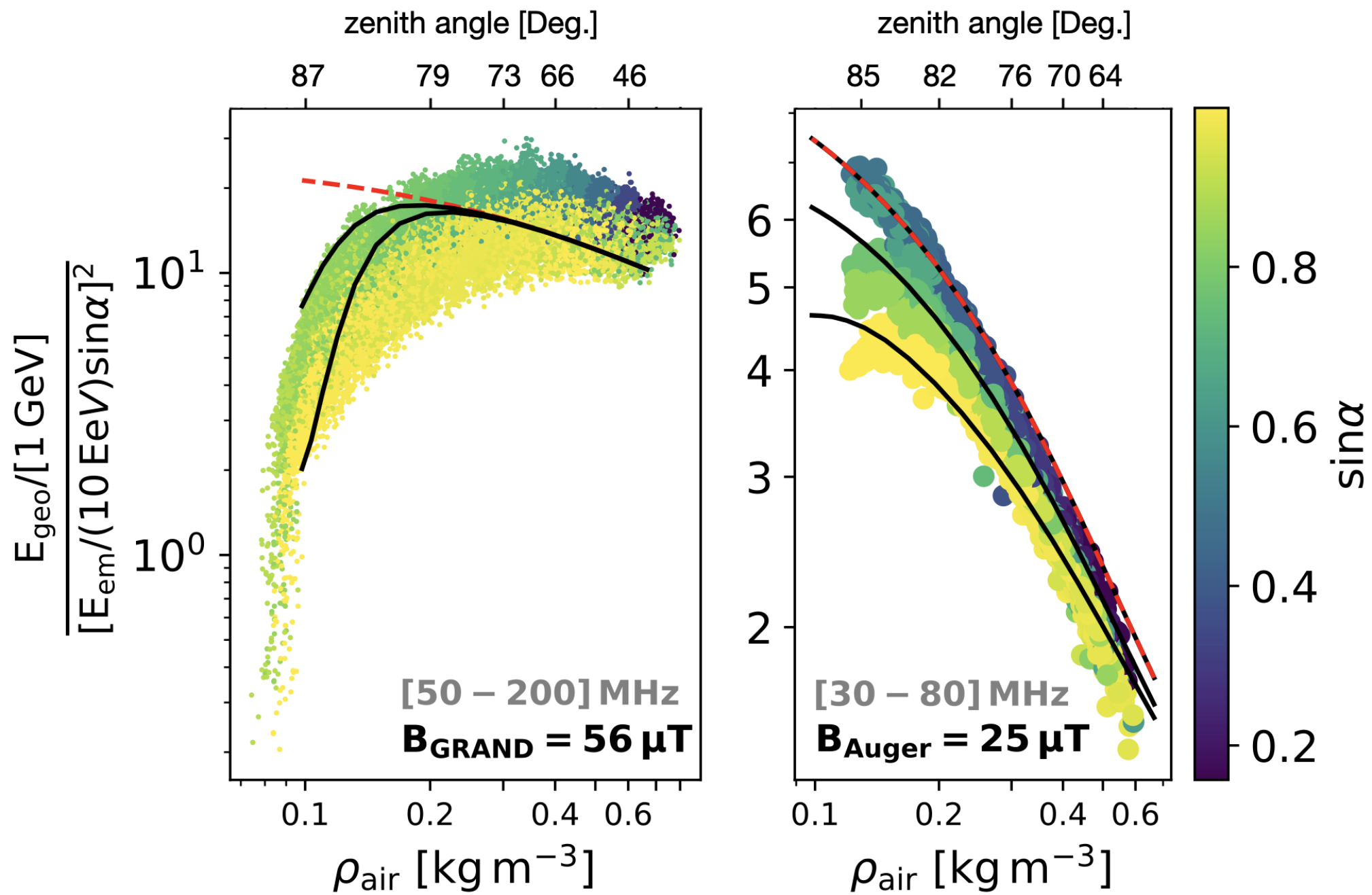}
\caption{{\it Left:} Projected component of the filtered electric field amplitude in the $50-200\,$MHz band for the $\mathbf{v} \times  (\mathbf{v} \times \mathbf{B})$ polarization, from  a ZHAireS simulation of a shower with zenith angle $\theta = 77^{\circ}$. The emission is strongest along the diagonal axis of the shower plane and follows a ``clover-leaf" pattern. 
 The same pattern is also observed with CoREAS simulations. {\it Right:}  Geomagnetic radiation energy as function of air density,  zenith angle and $\sin{\alpha}$ for GRAND magnetic field with ZHAireS simulations in the $50-200\,$MHz band (left panel, colored dots) and CoREAS simulations with Auger magnetic field in the $30-80\,$MHz band (right panel, colored dots). 
On both plots, the black lines follow our analytical modeling of the coherence effects discussed in Appendix~\ref{app:rad_energy}. For Auger, the three solid lines correspond to $\sin{\alpha} =1, 0.7, 0.3$ (bottom, middle and top). For GRAND, the two solid lines correspond to $\sin{\alpha} =1, 0.8$ (bottom and top). On both plots, the red dashed lines indicate the radiation energy expected if no coherence loss effects are modeled.}\label{fig:RadioSignatures} 


\end{figure*}

\section{\label{sec:numerical}  Microscopic simulations
\protect\\}
We compare the new emission signatures and regimes
presented in the previous section
 with predictions of numerical simulations, with the ZHAireS and CoREAS microscopic Monte-Carlo codes.

First, we investigated the polarization signatures by computing 
the $\mathbf{v\times (v \times B)}$ projection of the electric field amplitude in a plane perpendicular to the shower axis, for various shower parameters and GRAND-site Dunhuang magnetic field strength ($B_{\rm GRAND} = 56\, \mu T$).

Figure~\ref{fig:RadioSignatures} (left) typically displays the case of a shower with zenith angle $\theta = 77^{\circ}$. We observe a polarization pattern 
peaked along the diagonal axes of the ($\mathbf{v\times B}; \mathbf{v\times (v \times B)}$) plane that vanishes along the main axes, both for simulations with ZHAireS and CoREAS. This ``clover-leaf" polarization pattern is observed throughout the $(\nu,\rho)$ phase-space where geo-synchrotron can contribute significantly (above the orange lines in Fig.~\ref{fig:regimes}, right-hand panel). The pattern
cannot be described by 
the current macroscopic radio emission descriptions, neither by the transverse current nor by the Askaryan~\cite{Schr_der_2017}, 
emissions. It 
confirms the existence of a third type of emission that accounts for $\sim 10\%$ of the total radio signal amplitude. The geo-synchrotron component modeled in section~\ref{sec:synch} could explain this  pattern. Indeed, we expect that the individual emission of the electron/positron pairs will interfere constructively and destructively to give rise to a clover-leaf like polarization pattern with four maxima of emissions, as was discussed in~\cite{Huege2003, Huege_2005}.





We also evaluated the geomagnetic radiation energy predicted by ZHAireS and CoREAS as a function of air density, i.e., of zenith angle, 
following the method of Ref.~\cite{Glaser_2016,Chiche_2022}. 
We 
correct for any dependency on the azimuth and zenith angle or the primary energy by dividing by $E_{\rm em}^{2}\sin{\alpha}^{2}$ (with $E_{\rm em}$ the shower electromagnetic energy). We use a set of $\sim\, 10\,000$ ZHAireS and CoREAS showers with antennas on a star-shape layout, with zenith angle $\theta$ between $[40^{\circ}-87^{\circ}]$ for ZHAireS ($[65^{\circ}-85^{\circ}]$ for CoREAS), various azimuth angles $\phi$ and primary particle energy $\mathcal{E}$ between $[0.1-4]$\,EeV for ZHAireS ($[2.5-158]$\,EeV for CoREAS). 

The results are shown in the right-hand panel of Fig.~\ref{fig:RadioSignatures}, 
where we present the 
geomagnetic radiation energy as a function of air density and zenith angle for different magnetic field values. 
When going from high to low densities, the radiation energy first increases following a scaling in  $E^{\rm rad}_{\rm geo} \propto (1-p_0+p_0\exp{[p_1(\rho_{\rm Xmax} - \langle \rho \rangle) ]})^{2}$, as expected from~\cite{Glaser_2016,2023FelixJCAP...01..008S}, with $\langle \rho \rangle = 0.3\, \rm kg\, m^{-3}$, $p_0 \sim 0.5$ and $p_1 \sim -2.7\, \rm m^{3}/kg$ for Auger, $p_0 \sim 0.25$ and $p_1 \sim -1.8\, \rm m^{3}/kg$ for GRAND,  due to the different frequency band and also the stronger magnetic field amplitude resulting in a larger current at lower densities.
For GRAND parameters, however, it then drops by almost 1.5 orders of magnitude for air densities below $\rho_{\rm air} = 0.3\, \rm kg\, m^{-3}$ $(\theta \sim 70^{\circ})$. This radio emission cut-off is consistent with the incoherent regime described above and is qualitatively reproduced by our analytical modeling (Appendix~\ref{app:rad_energy}, black line). In contrast, for  Auger magnetic field, the geomagnetic radiation energy increases almost continuously when lowering the density and only a small slope change is observed at the lowest densities. 
The splitting at the lowest densities can be reproduced by our model (black lines) when considering different $\sin{\alpha}$ values, as the ``effective" magnetic field strength is given by $B\sin{\alpha}$. 
This result is consistent with the absence of a significant cut-off
observed at Auger. 

While the GRAND and Auger results are displayed for two different frequency bands, the main differences between both results arise from the change in the magnetic field, consistent with Eq.~\ref{eq:coherence}, which predicts scaling of the coherence ratio with $B^{2}$ while the scaling with frequency is linear. 

\section{Conclusions and perspectives 
\protect\\}\label{sec:level2}

Inclined cosmic-ray air showers develop higher in the atmosphere than  near-vertical ones and are expected to exhibit novel features that challenge our interpretation of radio emission. 
The lower air density should enhance magnetic deflections resulting in the emergence of a new polarization pattern, explainable with a  geo-synchrotron emission model, and a coherence loss due to the larger  shower lateral extension. 
Three different emission regimes are identified depending on the air density and magnetic field strength: (1) at the highest densities, the widely 
documented coherent transverse current emission, (2) at intermediate densities, a coherent emission with an additional component consistent with geo-synchrotron emission, (3)  at the lowest densities, incoherent emission. 

Our model is in agreement with two observational features predicted by both ZHAireS and CoREAS Monte-Carlo simulations, for strong enough magnetic field (typical strength of $B\sim 50\, \rm \mu T$): we observe a ``clover-leaf" polarization pattern in the $\mathbf{v} \times  (\mathbf{v} \times \mathbf{B})$ polarization and a cut-off in the emitted radiation energy of showers with zenith angle $\theta \gtrsim 70^{\circ}$. 

The new polarization pattern brings about a new paradigm in air shower radio emission as it breaks the current assumption of a radio signal made of transverse current and Askaryan emission only~\cite{Schr_der_2017, Huege2016, Schellart_2014, Aab_2014}. 
This implies that new reconstruction methods, accounting for this paradigm, are needed to prepare next-generation experiments that will target these very inclined air showers.

Our study also shows that the location of radio-detection experiments on the Earth can be chosen according to the geomagnetic field strength to enhance or suppress the cosmic-ray detection rate, depending on the scientific objectives. 
Finally, since neutrino-induced showers develop typically at higher densities than cosmic-ray ones~\cite{Chiche_2022}, no coherence loss nor clover-leaf pattern is expected in their emission, which could be valuable for cosmic-ray/neutrino discrimination. 

The coherence loss could serve to suppress the cosmic-ray detection rate and reduce the background contamination, while the clover-leaf emission should help discriminating between primaries via the polarization.

\begin{acknowledgments}
We thank Nikolaos Karastathis for his help in characterizing the polarization of the synchrotron emission. We also thank Olivier Martineau-Huynh, Washington R. Carvalho Jr, Pragati Mitra and Fabrice Mottez for the fruitful discussions on the radio cut-off. We thank the GRANDPa team for valuable discussions. More generally we thank the GRAND Analysis Forum for feedback on this work. We thank the referees for their thorough reading and insightful comments. This work was supported by the Deutsche Forschungsgemeinschaft (DFG, German Research Foundation) – 490843803 and by the PHC TOURNESOL program 48705ZF. KK is supported by the APACHE grant (ANR-16-CE31-0001) of the French Agence Nationale de la Recherche. C.Z. received financial support from the Helmholtz—OCPC Postdoc-Program. This work is supported by the European Research Council under the European Unions Horizon 2020 research and innovation program (No 805486 - K. D. de Vries). The authors acknowledge support by the state of Baden-Württemberg through bwHPC.

\end{acknowledgments}

\appendix

\section{Coherence effects from a particle population}\label{app:rad_energy}

To compute the observed radiation energy, we  assume that the total electric field  $\mathcal{E}_{\rm tot}$ measured at the observer location $x$ at time $t$, is given by Eq.~\ref{eq:Etot}.


We consider that all particles emit radio waves with the same wave number $k= 2 \pi/\lambda$ and with the same amplitude $\mathcal{E}_{0}$. The radio signal frequency $\nu = c/\lambda$ is set to $50\, \rm MHz$, as this value is covered by both Auger and GRAND frequency ranges and is representative of most of the power contained in those ranges. Because the geomagnetic deflections dominate the spatial distribution of the particles, we assume the source to extend along a one-dimensional line in the $\mathbf{v \times B}$ direction (i.e., we neglect the width of the particle distribution in the $\mathbf{v\times (v \times B)}$ direction). The scale of this spatial extent,  $l_{\rm lat}$ as defined in section~\ref{sec:Coherence}, can be derived from the extent of the particle energy content in the shower plane. CORSIKA simulations performed in~\cite{Guelfand2023arXiv231019612G} showed that $\sim 90\%$ of the particle energy is contained within $l_{\rm lat}$, making it a reasonable proxy to estimate the extent of the particle distribution. Although being a  simplification likely overestimating the width of the distribution, within that scale, we assume the particle distribution to be uniform and describe it with a top-hat function.

We then divide the shower lateral extent in $N_{\rm bins}$ of length $l_{\rm lat}/N_{\rm bins}$ and containing each $\bar{N} =N/N_{\rm bins}$ particles with the same energy, so that we get:
\begin{eqnarray}
     \mathcal{E}_{\rm tot} &=& \bar{N}\sum_{j=0}^{N_{\rm bins}}\mathcal{E}_0\cos{(kx - \omega t + \varphi_0 + \Delta \varphi_j)} \ ,
  \end{eqnarray}
where we defined $\Delta \varphi_j = \varphi_j - \varphi_0$, the phase difference between the $j^{\rm th}$ and the $0^{\rm th}$ bin, which we chose to be centered on $X_{\rm max}$.  The phase differences $\Delta \varphi_j$ depend on the shower lateral extent $l_{\rm lat}$ and are given by $\Delta \varphi_j = 2 \pi \delta_j /\lambda$, where $\lambda$ is the wavelength and the $\delta_j$ are the path differences between  the $X_{\rm max}$ - shower core  distance and the path from the center of the $j^{\rm th}$ bin and the shower core. Assuming that the refractive index $n$ is roughly equal to 1 along both paths and as we have $l_{\rm lat} \ll d_{\rm obs}$, we find: $\delta_j = j^{2}l^{2}_{\rm lat}/(2 N^{2}_{\rm bins} d_{\rm obs})$. 



Finally, the observed radiation energy is obtained by averaging over time the squared electric field at the observer location. The squared electric field has a periodicity $T=2\pi/\omega$, hence we average this quantity by numerical integration over one period of time:
\begin{eqnarray}\label{eq:Eobs_rad}
 E_{\rm rad}^{\rm obs} &=&  \left<|\mathcal{E}_{\rm tot}(t)|^{2}\right>_{T} = \frac{\omega}{2\pi}\int_{0}^{2\pi/\omega}|\mathcal{E}_{\rm tot}(t)|^{2}{\rm dt} \ .
\end{eqnarray}

The result is expressed as  $E_{\rm rad}^{\rm obs} = E_{\rm rad}^{\rm model}(\rho) f(\rho)$ where $E_{\rm rad}^{\rm model}(\rho)$, is the radiation energy expected from the existing theoretical descriptions of the radio emission and $f(\rho)$, is our correction due to the coherence effects, given by solving Eq.~\ref{eq:Eobs_rad}. We recall that we expect a scaling  in $E_{\rm rad}^{\rm model} \propto 1/(1-p_0+p_0\exp{[p_1(\rho_{\rm xmax} - \langle \rho \rangle) ]})^{2}$ from~\cite{Glaser_2016} which finally yields:  $E_{\rm rad}^{\rm obs} =A f(\rho)/(1-p_0+p_0\exp{[p_1(\rho_{\rm xmax} - \langle \rho \rangle) ]})^{2}$ where $A$, is a normalization factor that can be adjusted, $\langle \rho \rangle = 0.65\, \rm kg\, m^{-3}$ and $p_0$ and $p_1$ are fitting parameters.

\bibliography{biblio}

\end{document}